\documentclass[preprint,pra,showpacs,nofootinbib]{revtex4}
\usepackage{mathrsfs}
\usepackage{amsmath}

\def\be{\begin{equation}}
\def\ee{\end{equation}}
\def\ben{\begin{eqnarray}}
\def\een{\end{eqnarray}}

\begin{document}

\begin{flushright}
\vspace*{1cm}
\end{flushright}

\title{\bf Brownian motion of a charged test particle driven by \\ vacuum fluctuations near a dielectric half-space }
\author{Hongwei Yu\;$^{1,2}$, Xiangyun Fu\;$^{1,2}$ and Puxun Wu\;$^{3}$ }
\address
{$^1$Department of Physics and Institute of  Physics, Hunan Normal
University, Changsha, Hunan 410081, China
\\
$^2$Key Laboratory of Low Dimensional Quantum Structures and Quantum
Control of Ministry of Education, Hunan Normal University ,
Changsha, Hunan 410081, P.R. China
 \\
$^3$Department of Physics and Tsinghua Center for Astrophysics,
Tsinghua University, Beijing 100084, China
}


\begin{abstract}

 We study the Brownian motion of a charged
test particle driven  by quantum electromagnetic  fluctuations in
the vacuum region near a non-dispersive and non-absorbing dielectric
half-space and calculate the mean squared fluctuations in the
velocity of the test particle.  Our results show that a nonzero
susceptibility of the dielectrics has its imprints on the velocity
dispersions of the test particles.  The most noteworthy feature in
sharp contrast to the case of an idealized perfectly conducting
interface is that the velocity dispersions in the parallel
directions are no longer negative and does not die off in time,
suggesting that the potentially problematic negativeness of the
dispersions in those directions in the case of perfect conductors is
just a result of our idealization and does not occur for real
material boundaries.

\end{abstract}
\pacs{05.40.Jc, 12.20.Ds, 03.70.+k}

\maketitle \baselineskip=14pt


\section{Introduction }


 A fundamental feature to be expected of any quantized field is the quantum vacuum fluctuations.
 Although the effects of electromagnetic vacuum fluctuations upon a test particle may be
 unobservable in free space in quantum electrodynamics,
it is well-known that changes in the vacuum fluctuations induced by
the presence of boundaries can produce novel observable effects.
Typical examples are the Casimir effect~\cite{Cas}, the Lamb shift
and modified spontaneous emission rates and energy shifts for an
atom near a reflecting surface~\cite{MJH,Yu05,Yu06}.

Since there always exist quantum electromagnetic vacuum
fluctuations, one would expect that  test particles under the
influence of these quantum field fluctuations will no longer move on
the classical trajectories, but undergo random motion around a mean
path. Recently, this kind of random (Brownian) motion, as opposed to
that driven by classical or thermal fluctuations~\cite{Einstein},
has been investigated for a charged test particle near a perfectly
reflecting plane and between parallel
plates~\cite{HFord,HCh,HChw,SWu}  and the effects have been
calculated of the modified electromagnetic vacuum fluctuations due
to the presence of the boundary upon the motion of a charged test
particle. In particular, it has been shown that the mean squared
fluctuations in the normal velocity and position of the test
particle can be associated with an effective temperature which is
possibly experimentally accessible~\cite{HFord}.  To date all such
field-theoretical calculations have been done for boundaries that
are idealized as perfect conductors. However, it is obvious that no
real material can ever really be perfectly conducting. Therefore it
remains interesting to see what happens when the perfect conductors
are replaced by imperfect conductors with finite refractivity.

 In the present paper, we will examine a situation where a single
 interface is located at $z=0$, and the region to the left of this
 interface ($z<0$ ) is filled with a non-dispersive and non-absorbing
 dielectric while the region to the right ($z>0$) is vacuum. We will
 compute the effects of quantum electromagnetic vacuum fluctuations
 upon the motion of a charged test particle in the vacuum region.
 Since the boundary now is no longer a perfect conductor, one needs
 a quantization scheme for the electromagnetic fields which takes into
 account the evanescent waves in addition to the incident and
 reflected ones. Fortunately, such a quantization procedure has been
 given by Carniglia and Mandel~\cite{carm}


\section{The orthonormal eigenmodes and the two-point function}


For simplicity, we model the imperfect conductor by a uniform
nondispersive and nonabsorbing dielectric half-space in the region
$z < 0$. Thus, the dielectric medium is characterized by a single
parameter, its refractive index, which is real and
frequency-independent. The refractive index can then be written as
\begin{eqnarray}
n({\bf r})=1+\theta (-z)\,(\epsilon-1)\;.
\end{eqnarray}
Here $\epsilon$ is the dielectric constant of the medium which can
be expressed, in term of the dielectric susceptibility $\chi$, as
$\epsilon=1+\chi$ and $\theta (z)$ is the step function.

Let us now consider the motion of a charged test particle subject to
quantum electromagnetic vacuum fluctuations in the vacuum region(
$z>0$) with a dielectric half-space. In the limit of small
velocities and assuming that the particle is initially at rest and
has a charge to mass ratio of $ e/m$,  the mean squared speed in the
$i$-direction can be written as (no sum on $i$)
\begin{eqnarray}
\langle{\Delta v_i^2}\rangle&=&{e^2\over m^2}\;\int_0^t\;\int_0^t\;
\langle{E}_i({\mathbf
x},t_1)\;{E}_i({\mathbf x},t_2)\rangle_R\,dt_1\,dt_2\nonumber\\
&=&{4\pi\alpha\over m^2}\;\int_0^t\;\int_0^t\; \langle{E}_i({\mathbf
x},t_1)\;{E}_i({\mathbf x},t_2)\rangle_R\,dt_1\,dt_2
 \; ,   \label{eq:lang2}
\end{eqnarray}
where $\alpha$ is the fine-structure constant and $
\langle{E}_i({\mathbf x},t_1)\;{E}_i({\mathbf x},t_2)\rangle_R $ is
the renormalized electric field two-point function obtained by
subtracting the boundary-independent Minkowski vacuum term.  We
have, for simplicity, assumed that the distance does not change
significantly in a time $t$, so that it can be treated approximately
as a constant. If there is a classical, nonfluctuating field in
addition to the fluctuating quantum field, then Eq.~(\ref{eq:lang2})
describes the velocity fluctuations around the mean trajectory
caused by the classical field. Note that when the initial velocity
does not vanishes, one
 has to also consider the influence of fluctuating magnetic fields
 on the velocity dispersions of the test particles. However, it has
 been shown that this influence is, in general, of
the higher order than that caused by fluctuating electric fields and
is thus negligible \cite{TY}.

In order to compute the dispersions of the random motion of the test
particle, we need the renormalized electric field two-point
function, $\langle{E}_i({\mathbf x},t_1)\;{E}_i({\mathbf
x},t_2)\rangle_R $. Fortunately, this function can be calculated
using the  orthonormal eigenmodes  and the quantization procedure
with the evanescent waves taken into account,  as given by Camiglia
and Mandel in~\cite{carm}. To calculate the renormalized electric
field two-point function, let $x_T=(x,y)$ denote directions
tangential to the interface and $k_T$ be Fourier transform variable
associated with it.  If the wavenumber in the $z$-direction is $k$
in the vacuum and $k_D$ in the dielectric, then we
\begin{eqnarray}
k_D^2+k_T^2&=&\epsilon\,\omega^2 \;\;\;\;{\text {for}}\;\;\;\; z<0 \\
\nonumber k^2+k_T^2&=&\omega^2 \;\;\;\,\;\;{\text {for}}
\;\;\;\;z>0.
\end{eqnarray}
 Since we are interested in the random motion of test particles in the vacuum region, we
 will
only need the electric modes on the vacuum side of the interface.

Following Ref.\,\cite{carm,adam},
one can show that
\begin{eqnarray}
\langle{{E}_{z}}({{\bf x}},t'),{{E}_{z}}({{\bf
x}},t'')\rangle&=&{1\over (2\pi)^3}\int_{k_D>0}d^3k_D{k_T^2
\over{2\,\omega}}\bigg({2k_D\over{k_D+\epsilon k}}\bigg)
\bigg({2k_D\over{k_D+\epsilon k}}\bigg)^{\ast}e^{i(k-k^{\ast})z}
 e^{i\omega \Delta t}\nonumber\\&&\quad+{1\over
(2\pi)^3}\int_{k>0}d^3k{k_T^2
\over{2\,\omega}}\bigg(e^{-ikz}+{\epsilon k-k_D\over{\epsilon
k+k_D}}e^{ikz}\bigg) \nonumber\\ &&
\quad\quad\quad\quad\quad\quad\times\bigg(e^{-ikz}+{\epsilon
k-k_D\over{\epsilon k+k_D}}e^{ikz}\bigg)^{\ast}e^{i\omega \Delta
t}\;,
\end{eqnarray}
and
\begin{eqnarray}
&&\langle{{E}_{x}}({{\bf x}},t'),{{E}_{x}}({{\bf
x}},t'')\rangle=\langle{{E}_{y}}({{\bf x}},t'),{{E}_{y}}({{\bf
x}},t'')\rangle \nonumber\\
&&\;={1\over (2\pi)^3}\int_{k_D>0}d^3k_D\bigg[{k^2
\over{4\,\omega}}\bigg({2k_D\over{k_D+\epsilon k}}\bigg)
\bigg({2k_D\over{k_D+\epsilon k}}\bigg)^{\ast}+{1
\over{4\epsilon}}\bigg({2k_D\over{k_D+ k}}\bigg)
\bigg({2k_D\over{k_D+
k}}\bigg)^{\ast}\bigg]e^{i(k-k^{\ast})z}e^{i\omega \Delta t}
\nonumber\\
&&\;\;+{1\over (2\pi)^3}\int_{k>0}d^3k\bigg[{k^2
\over{4\,\omega}}\bigg(e^{-ikz}-{\epsilon k-k_D\over{\epsilon
k+k_D}}e^{ikz}\bigg)\bigg(e^{-ikz}-{\epsilon k-k_D\over{\epsilon
k+k_D}}e^{ikz}\bigg)^{\ast}\nonumber\\
&&\quad\quad\quad\quad\quad\quad\quad+{1 \over{4}}\bigg(e^{-ikz}+{
k-k_D\over{ k+k_D}}e^{ikz}\bigg) (e^{-ikz}+{ k-k_D\over{
k+k_D}}e^{ikz}\bigg)^{\ast}\bigg]e^{i\omega \Delta t}.
\end{eqnarray}
where $i\Delta t =t'-t''$. Adopting the method of Ref.~\cite{adam},
the two-point functions can be rewritten, after performing the angle
integration, as
\begin{eqnarray}
\langle{{E}_{z}}({{\bf x}},t'),{{E}_{z}}({{\bf
x}},t'')\rangle&=&{1\over
(2\pi)^2}\int_{0}^{\infty}d\omega\int_{0}^{1}d\xi\,\omega^3(1-\xi^2)\bigg(1+{\epsilon\,\xi-\sqrt{\chi+\xi^2}
\over \epsilon\,\xi+\sqrt{\chi+\xi^2}}\cos2\omega\xi z\bigg)
e^{i\omega \Delta t}\nonumber\\&&+{1\over
(2\pi)^2}\int_{0}^{\infty}d\omega\int_{0}^{1}d\xi\,\omega^3(1+\chi\xi^2)
{2\epsilon {\sqrt\chi} \xi \sqrt{(1-\xi^2)} \over
1+(\epsilon^2-1)\xi^2}e^{-2\omega\sqrt{\chi}\,\xi z} e^{i\omega
\Delta t}\; ,\nonumber\\
\end{eqnarray}
and
\begin{eqnarray}
&&\langle{{E}_{x}}({{\bf x}},t'),{{E}_{x}}({{\bf
x}},t'')\rangle=\langle{{E}_{y}}({{\bf x}},t'),{{E}_{y}}({{\bf
x}},t'')\rangle\nonumber\\&&\,={1\over
8\pi^2}\int_{0}^{\infty}d\omega\int_{0}^{1}d\xi\,\omega^3\bigg[\,\xi^2\bigg(1-{\epsilon\,\xi-\sqrt{\chi+\xi^2}
\over
\epsilon\,\xi+\sqrt{\chi+\xi^2}}\bigg)+\bigg(1+{\,\xi-\sqrt{\chi+\xi^2}
\over \,\xi+\sqrt{\chi+\xi^2}}\bigg)\bigg]\cos2\omega\xi z\;
e^{i\omega \Delta t}\nonumber\\&&\quad\quad\quad+{1\over
8\pi^2}\int_{0}^{\infty}d\omega\int_{0}^{1}d\xi\,\omega^3\,\bigg(
{2\xi^3\,\epsilon\chi^{3/2}\sqrt{1-\xi^2} \over
1+(\epsilon^2-1)\xi^2}+ {2\,\xi\chi^{1/2}\sqrt{1-\xi^2}
}\bigg)e^{-2\omega\sqrt{\chi}\,\xi z} e^{i\omega \Delta t}\;
.\nonumber\\
\end{eqnarray}
Here, the first integral represents the contribution of plane waves
and the second  that of evanescent waves. Integrating over $\omega$
gives
\begin{eqnarray}
\langle{{E}_{z}}({{\bf x}},t'),{{E}_{z}}({{\bf
x}},t'')\rangle&=&{1\over (2\pi)^2}\int_{0}^{1}d\,\xi\bigg[6{1-\xi^2
\over \Delta t^4}+6(1-\xi^2){\epsilon\,\xi-\sqrt{\chi+\xi^2} \over
\epsilon\,\xi+\sqrt{\chi+\xi^2}}\nonumber\\
&&\quad\quad\quad\quad\quad\quad\times\frac{16z^4\xi^4+24z^2\xi^2\Delta
t^2+\Delta t^4}{(4z^2\xi^2-\Delta
t^2)^4}\nonumber\\
&&\quad\quad\quad\quad\quad+\frac{12\epsilon\sqrt{\chi}
\xi(1+\chi\xi^2)\sqrt{1-\xi^2}}{[1+(\epsilon^2-1)\xi^2](2z\sqrt{\chi}\xi-i\Delta
t)^4}\bigg]\;,
\end{eqnarray}
and
\begin{eqnarray}
&&\langle{{E}_{x}}({{\bf x}},t'),{{E}_{x}}({{\bf
x}},t'')\rangle=\langle{{E}_{y}}({{\bf x}},t'),{{E}_{y}}({{\bf
x}},t'')\rangle\nonumber\\&&\quad={1\over
8\pi^2}\int_{0}^{1}d\,\xi\bigg[{6(1+\xi^2 )\over \Delta
t^4}+\bigg({6(\xi-\sqrt{\chi+\xi^2}) \over
\xi+\sqrt{\chi+\xi^2}}-6\xi^2{\epsilon\xi-\sqrt{\chi+\xi^2} \over
\epsilon\xi+\sqrt{\chi+\xi^2}}\,\bigg)\nonumber\\
&&\quad\quad\quad\quad\quad\quad\quad\quad\times
\frac{16z^4\xi^4+24z^2\xi^2\Delta t^2+\Delta t^4}{(4z^2\xi^2-\Delta
t^2)^4} \nonumber\\
&&\quad\quad\quad\quad\quad\quad+\frac{12\epsilon
{\chi}^{3/2}\xi^3\sqrt{1-\xi^2}}{[1+(\epsilon^2-1)\xi^2](2z\sqrt{\chi}\xi-i\Delta
t)^4}+\frac{12
\chi^{1/2}\xi\sqrt{1-\xi^2}}{(2z\sqrt{\chi}\xi-i\Delta
t)^4}\bigg]\;.
\end{eqnarray}
The two-point functions can be expressed as the sum of Minkowski
vacuum term and a correction term due to the dielectric half-space
boundary. The first term, after being integrated over $\xi$, is ${1/
({\pi^2\Delta t^4}})$. So, it is the Minkowski vacuum term and shall
be dropped because it is not expected to produce any observable
consequences. Then we  obtain  the renormalized two-point function
\begin{eqnarray}
&&\langle{{E}_{z}}({{\bf x}},t'),{{E}_{z}}({{\bf
x}},t'')\rangle_R\nonumber\\&&\quad\,={1\over
(2\pi)^2}\int_{0}^{1}d\,\xi\bigg[6(1-\xi^2){\epsilon\,\xi-\sqrt{\chi+\xi^2}
\over
\epsilon\,\xi+\sqrt{\chi+\xi^2}}\frac{16z^4\xi^4+24z^2\xi^2\Delta
t^2+\Delta t^4}{(4z^2\xi^2-\Delta
t^2)^4}\nonumber\\
&&\quad\quad\quad\quad\quad\quad\quad\quad\quad+\frac{12\epsilon\sqrt{\chi}
\xi(1+\chi\xi^2)\sqrt{1-\xi^2}}{[1+(\epsilon^2-1)\xi^2](2z\sqrt{\chi}\xi-i\Delta
t)^4}\bigg]\;,
\end{eqnarray}
and
\begin{eqnarray}
&&\langle{{E}_{x}}({{\bf x}},t'),{{E}_{x}}({{\bf
x}},t'')\rangle_R=\langle{{E}_{y}}({{\bf x}},t'),{{E}_{y}}({{\bf
x}},t'')\rangle_R\nonumber\\&&\quad={1\over
8\pi^2}\int_{0}^{1}d\,\xi\bigg[\bigg({6(\xi-\sqrt{\chi+\xi^2}) \over
\xi+\sqrt{\chi+\xi^2}}-6\xi^2{\epsilon\xi-\sqrt{\chi+\xi^2} \over
\epsilon\xi+\sqrt{\chi+\xi^2}}\,\bigg)
\frac{16z^4\xi^4+24z^2\xi^2\Delta t^2+\Delta t^4}{(4z^2\xi^2-\Delta
t^2)^4} \nonumber\\
&&\quad\quad\quad\quad\quad\quad\quad+\frac{12\epsilon
{\chi}^{3/2}\xi^3\sqrt{1-\xi^2}}{[1+(\epsilon^2-1)\xi^2](2z\sqrt{\chi}\xi-i\Delta
t)^4}+\frac{12
\chi^{1/2}\xi\sqrt{1-\xi^2}}{(2z\sqrt{\chi}\xi-i\Delta
t)^4}\bigg]\;.
\end{eqnarray}


\section{ the Brownian motion of the test particle}

With the renormalized two-point functions found, we now start to
compute the velocity dispersions using Eq.~(\ref{eq:lang2}).
Although the integrations can be performed into closed form, the
general results are very tedious and not very illuminating. We will
not give them here, but, instead, we will analyze some special cases
of interest. The first is what happens to the velocity dispersions
at later times when the dielectric susceptibility $\chi$ is very
large (the dielectric deviates significantly from a perfect
conductor), i.e., we will analyze the behavior of both
$\langle\,\Delta v_z^2\,\rangle$ and $\langle\,\Delta
v_x^2\,\rangle$ in the limit of $\chi\rightarrow \infty $ and $t\gg
z$. In this limit, we have

\begin{eqnarray}
\langle\,\Delta v_z^2\,\rangle&\approx&{e^2\over 4\pi^2
m^2}\;\int_0^t\,dt'\;\int_0^t d
t''\;\int_{0}^{1}d\,\xi\,\bigg[6(1-\xi^2){\epsilon\,\xi-\sqrt{\chi+\xi^2}
\over
\epsilon\,\xi+\sqrt{\chi+\xi^2}}\times\nonumber\\
&&\quad\quad\frac{16z^4\xi^4+24z^2\xi^2\Delta t^2+\Delta
t^4}{(4z^2\xi^2-\Delta t^2)^4}+\frac{12\epsilon\sqrt{\chi}
\xi(1+\chi\xi^2)\sqrt{1-\xi^2}}{[1+(\epsilon^2-1)\xi^2](2z\sqrt{\chi}\xi-i\Delta
t)^4}\bigg]\nonumber\\
&\approx&\;{e^2\over 4 \pi^2 m^2}{1\over z^2} + {e^2\over 3\pi^2
m^2}{1\over t^2} +{e^2\over 4 \pi^2 m^2 }\biggl({2\ln(4\chi)\over
 t^2\sqrt{\chi}}+{1+\ln(4\chi)\over
 2 z^2\sqrt{\chi}}\biggr)\nonumber\\
&\approx&\; {e^2\over 4 \pi^2 m^2}{1\over
z^2}\bigg(1+{1+\ln(4\chi)\over
 2\sqrt{\chi}}\bigg) + {e^2\over
3\pi^2 m^2}{1\over t^2}\bigg(1+{3\ln(4\chi)\over
 2\sqrt{\chi}}\bigg)\;.
 \label{Vz}
\end{eqnarray}
In the above expression, one sees clearly that the first two terms
are just the result for a perfectly conducting interface (refer to
Ref.~\cite{HFord}), while the $\chi$ dependent terms are corrections
induced by finite refractivity.

Similarly, one has, for the velocity dispersions in the directions
parallel to the interface, in the limit $\chi\rightarrow \infty $
and $t\gg z$
\begin{eqnarray}
\langle\Delta v_x^2\rangle&=&\langle\Delta
v_y^2\rangle\approx{e^2\over 4 \pi^2 m^2}{1\over
z^2}\bigg({\ln(4\chi) -1)\over
 2\sqrt{\chi}}\bigg)-{e^2\over
3\pi^2 m^2}\bigg(1-{1\over \sqrt{\chi}}\bigg){1\over t^2}\,.
\label{Vx}
\end{eqnarray}
Once again, we can see that when $\chi$ approaches infinity, we
recover the result found in Ref.~\cite{HFord} for the vacuum
half-space outside a perfectly conducting plane. This is what one
would expect since perfect conductors are formally identified with
dielectrics of infinite susceptibility. However, the approach to
this limit in terms of the velocity dispersions of charged test
particles under Brownian motion driven by vacuum fluctuations is
slow. Take the dispersion in the normal direction for an example, we
need $\chi \approx 2632$ for $\langle\,\Delta v_z^2\,\rangle$ to be
within $10 \% $ of its limiting value, and $\chi \approx 14288 $ to
be within $5 \% $. Eq.~(\ref{Vz}) and Eq.~(\ref{Vx}) show that the
corrections induced by finite refractivity are positive and hence
the velocity dispersions of the test particle outside real materials
would be larger than those outside an idealized perfect conductor.
The most noteworthy feature in sharp contrast to an idealized
perfectly conducting interface~\cite{HFord} is that the velocity
dispersions in the parallel directions, $\langle\Delta v_x^2\rangle$
and $\langle \Delta v_y^2\rangle$, are no longer negative and does
not die off in time. Therefore, it is not a transient effect any
more. Finally, let us note that the velocity dispersions all
approach nonzero constant values at late times in contrast to
Brownian motion due to thermal noise, where dissipation is needed
for $\langle \Delta v_i^2\rangle$ to be bounded at late times.. The
fact they do not continue to grow in time can be understood as a
consequence of energy conservation and they may be absorbed into a
renormalization of the mass of the test particles~\cite{ICGA}.

For the sake of completeness, let us now  also give the results for
the velocity dispersions when the dielectric susceptibility $\chi$
is vanishingly small (corresponding to vacuum). In this case, we
have
\begin{eqnarray}
\langle\,\Delta v_z^2\,\rangle&\approx&{e^2\,\chi\over 4\pi^2
m^2z^2}\bigg({t^2+12z^2\over 12t^2z^2}+{(t^4 - 8 z^4)\over 16 t^3z^3
}\ln\biggl( {2z+t\over
 2z-t}\biggr)^2\,\bigg)\nonumber\\&\approx&
 \left\{
 \begin{aligned}
 &{e^2\over 4\pi^2
m^2z^2}\bigg({7  \over 12  z^2} + {5 \over
 3 t^2}\bigg)\,\chi\,,\quad\quad t\rightarrow \infty\,,\cr
 & {9\, e^2 t^2 \chi \over 320 m^2 \pi^2 z^4}\,,\quad\quad  t\rightarrow
 0\,,\cr
 \end{aligned} \right.
\end{eqnarray}
and
\begin{eqnarray}
\langle\,\Delta v_x^2\,\rangle=\langle\,\Delta
v_y^2\,\rangle&\approx&{e^2\over 4\pi^2 m^2z^2}\bigg(-{t^4 - 4 t^2
z^2 + 24 z^4 \over
   z^2 (t^4 - 4 t^2 z^2)} +{(t^4 - 8 z^4)\over 32 t^3z^3
}\ln\biggl( {2z+t\over
 2z-t}\biggr)^2\,\bigg)\nonumber\\&\approx&
 \left\{
 \begin{aligned}
 &{e^2\over 4\pi^2
m^2z^2}\bigg({1\over 6 z^2} + {1\over
 3 t^2}\bigg)\,\chi\,,\quad\quad (t\rightarrow \infty)\,,\cr
& {7\, e^2 t^2 \chi \over 320 m^2 \pi^2 z^4}\,,\quad\quad
t\rightarrow
 0\,,\cr
 \end{aligned} \right.
\end{eqnarray}
As expected, all the above values of the velocity dispersions tend
to zero (linearly) as $\chi \rightarrow 0$. The same as the case
when $\chi \rightarrow \infty$, here the velocity dispersions in the
parallel directions are also positive. This suggests that the
negative values of velocity dispersions in the parallel directions
obtained in Ref.~\cite{HFord} is a result of idealization of the
boundary as perfectly conducting. In reality, a real conductor would
not be well approximated by a perfect conductor within atomic
distances and its plasma frequency.  Therefore, the potentially
problematic negativeness of the dispersions in those directions in
the case of perfect conductors  does not occur for real material
boundaries. It should be pointed out that the above expressions are
singular at $t=2z$. This corresponds to a time interval equal to the
round-trip light travel time between the particle and the interface.
Presumably, this might be a result of our assumption of a rigid
plane boundary, and would thus be smeared out in a more realistic
treatment, where fluctuations in the position of the interface are
taken into account.

\section{Summary}

 In summary, we have studied the Brownian motion of a charged
test particle driven  by quantum electromagnetic  fluctuations in
the vacuum region near a non-dispersive and non-absorbing dielectric
half-space and calculated the mean squared fluctuations in the
velocity of the test particle.  Our results show that a nonzero
susceptibility of the dielectrics has its imprints on the velocity
dispersions of the test particles. When the susceptibility is small,
the dependence of the velocity dispersions on it is linear. However,
when the susceptibility is extremely large, the dependence is a
rather complicated function involving logarithmic and square root.
In comparison with the case where the interface is an idealized
perfectly conducting boundary (infinite susceptibility),
 the dispersions of the test particle near a
dielectric half-space are larger than near a perfect conducting
boundary and they approach that of a perfect conducting boundary
very slowly as susceptibility increases. The most noteworthy feature
in sharp contrast to the case of an idealized perfectly conducting
interface~\cite{HFord} is that the velocity dispersions in the
parallel directions are no longer negative and does not die off in
time. Therefore, the potentially problematic negativeness of the
dispersions in those directions in the case of perfect conductors is
just a result of our idealization and does not occur for real
material boundaries.

\begin{acknowledgments}
 This work was supported in part by the National Natural Science
Foundation of China under Grants No.10575035, 10775050, the SRFDP
under Grant No. 20070542002, and the Programme for the Key
Discipline in Hunan Province.
\end{acknowledgments}

\end{document}